\documentclass[preprint,preprintnumbers, prd, floatfix, superscriptaddress,nofootinbib] {revtex4-1}
\usepackage{epsfig}
\usepackage{subfigure}
\usepackage{dcolumn}
\usepackage{bm}
\usepackage[usenames ,dvipsnames]{xcolor}
\usepackage{slashed}
\usepackage{graphicx,color}
\usepackage{hyperref}

\begin{document}
\title{Charmed $\Lambda_c^+$ baryon decays into light scalar mesons\\
in the topological $SU(3)_f$ framework}

\author{Y.~L.~Wang}
\email{ylwang0726@163.com}
\affiliation{School of Physics and Electronic Engineering, Shanxi Normal University,
Taiyuan 030031, China}

\author{Y.~K.~Hsiao}
\email{yukuohsiao@gmail.com}
\affiliation{School of Physics and Electronic Engineering, Shanxi Normal University,
Taiyuan 030031, China}

\date{\today}

\begin{abstract}
Using the topological-diagram approach based on $SU(3)$ flavor symmetry,
we investigate two-body $\Lambda_c^+\to {\bf B}S$ decays,
where ${\bf B}$ denotes the final-state baryon and $S$ refers to a light scalar meson,
such as $f_0/f_0(980)$, $a_0/a_0(980)$, $\sigma_0/f_0(500)$, or $\kappa/K_0^*(700)$. 
Our analysis indicates that interpreting the light scalar mesons as 
tetraquark states provides a more consistent description of 
the currently available experimental data. In particular, 
this framework naturally accommodates the experimentally observed branching fraction
${\cal B}(\Lambda_c^+ \to \Lambda a_0^+)$, which exceeds predictions
based solely on long-distance effects by an order of magnitude.
Within the tetraquark scenario, we predict
${\cal B}(\Lambda_c^+\to \Sigma^+ f_0)=(4.9\pm 1.9)\times 10^{-2}$ and
${\cal B}(\Lambda_c^+\to p f_0)=(3.6\pm 1.4)\times 10^{-3}$. 
Owing to $f_0-\sigma_0$ mixing, ${\cal B}(\Lambda_c^+\to \Sigma^+ \sigma_0)$ and 
${\cal B}(\Lambda_c^+\to p\sigma_0)$ are suppressed to 
the levels of $1\times 10^{-3}$ and $5\times 10^{-5}$, respectively.
These modes therefore provide sensitive probes of 
the internal structure of the light scalar mesons. 
More generally, the remaining branching ratios of $\Lambda_c^+ \to {\bf B} S$
are found to be comparable to those of $\Lambda_c^+ \to {\bf B}M$, where 
$M$ denotes a pseudoscalar meson. Their predicted sizes suggest that 
these decay modes should be accessible to ongoing and 
near-future experimental studies at BESIII, Belle II, and LHCb.
\end{abstract}

\maketitle
\section{introduction}
Exotic multi-quark bound states, such as tetraquarks and pentaquarks,
are predicted by the quark model to coexist with conventional mesons ($q\bar q$)
and baryons ($qqq$)~\cite{Gell-Mann:1964ewy,zweig}. However,
the existence of charmless exotic states remains inconclusive.
Among the most promising candidates for such states are the light scalar mesons ($S$):
$f_0/f_0(980)$, $a_0/a_0(980)$, $\sigma_0/f_0(500)$, and $\kappa/K_0^*(700)$,
with the masses below 1 GeV.
These states are often proposed to possess a tetraquark structure,
either as tightly bound (compact) $q^2\bar q^2$ states with gluon exchange~\cite{Jaffe:1976ig,
Jaffe:1976ih,Close:2002zu,Pelaez:2003dy,Maiani:2004uc,Amsler:2004ps,
Jaffe:2004ph,Achasov:2005hm,tHooft:2008rus,Fariborz:2009cq,
Weinberg:2013cfa,Agaev:2018fvz,Hsiao:2023qtk}, or as loosely bound meson-meson molecular states
held together by residual strong force~\cite{Weinstein:1990gu,Branz:2007xp,Baru:2003qq,
Dai:2012kf,Dai:2014lza,Sekihara:2014qxa,Yao:2020bxx,Wang:2022vga}.
On the other hand, a conventional p-wave $q\bar q$ interpretation has not yet been definitively excluded~\cite{Anisovich:2000wb,Bediaga:2003zh,Aliev:2007uu,Colangelo:2010bg,
Shi:2015kha,Soni:2020sgn,Klempt:2021nuf}.
To differentiate between these competing scenarios,
extensive studies have been carried out in $B$-meson, $D$-meson,
and $b$-baryon decays~\cite{pdg}. In contrast,
investigations of charmed baryon (${\bf B}_c$) decays remain relatively  scarce.

The first branching fraction of a ${\bf B}_c \to {\bf B} S$ decay,
${\cal B}_{\rm ex}(pf_0)\equiv{\cal B}(\Lambda_c^+ \to p f_0)= (3.5 \pm 2.3) \times 10^{-3}$,
was reported in 1990~\cite{Barlag:1990yv,pdg}. Owing to its large uncertainty, however,
the result left doubts about the feasibility of observing light scalar mesons
in charmed baryon decays at measurable rates.
This finding is broadly compatible with theoretical expectations, which place
${\cal B}({\bf B}_c \to {\bf B} S)$ in the range $10^{-5}-10^{-4}$~\cite{Sharma:2009zze},
near the edge of experimental sensitivity.
More recently, interest was revived by the suggestion in Ref.~\cite{Yu:2020vlt}
of a resonant contribution in $\Lambda_c^+ \to \Lambda (a_0^+ \to) \eta \pi^+$,
motivated by a faint structure observed in the Dalitz plot of
the Cabibbo-allowed decay $\Lambda_c^+ \to \Lambda \eta \pi^+$
reported by Belle~\cite{Lee:2020xoz}. Building on this,
BESIII has recently measured the relevant channel, finding
${\cal B}_{\rm ex}(\Lambda a_0^+) \equiv {\cal B}(\Lambda_c^+ \to \Lambda a_0^+)
= (1.23 \pm 0.21) \times 10^{-2}$~\cite{BESIII:2024mbf}.

While ${\cal B}_{\rm ex}(\Lambda a_0^+)$ reaches the order of $10^{-2}$,
far above the theoretical expectation~\cite{Yu:2020vlt},
one could in principle enhance ${\cal B}_{\rm th}(\Lambda a_0^+)$ by an order of magnitude
to reproduce the measurement. Such an enhancement, however,
would require an arbitrary tuning of the cut-off parameter in the triangle-loop calculation~\cite{Hsiao:2019ait,Yu:2021euw,Hsiao:2024szt,Yu:2020vlt}.
Other long-distance effects may also contribute to
${\bf B}_c \to {\bf B} S$~\cite{Sharma:2009zze, Xie:2016evi, Wang:2020pem,
Feng:2020jvp, Wang:2022nac}. In particular, both the pole model~\cite{Sharma:2009zze}
and the rescattering mechanism~\cite{Wang:2022nac}
predict ${\cal B}(\Lambda_c^+ \to \Lambda a_0^+)$ below $1\times 10^{-3}$~\cite{BESIII:2024mbf},
well short of the observed value~\cite{BESIII:2024mbf}.

As a possible resolution, we propose that short-distance effects dominate
${\bf B}_c \to {\bf B} S$ decays, in close analogy with the well-established short-distance dominance
in ${\bf B}_c \to {\bf B} M$ processes, where $M$ denotes a pseudo-scalar meson.
This interpretation is further supported by several experimental comparisons. First,
${\cal B}_{\rm ex}(pf_0)$ is of the same order of magnitude as ${\cal B}(\Lambda_c^+ \to p \eta')$,
and both $f_0$ and $\eta'$ are expected to contain sizable $s\bar{s}$ components
due to the analogous $\sigma_0-f_0$ and $\eta-\eta'$ mixings, respectively.
Second, we find that ${\cal B}_{\rm ex}(\Lambda a_0^+)\simeq {\cal B}(\Lambda_c^+ \to \Lambda \pi^+)$,
where $a_0$ and $\pi$ are the isospin triplets in the scalar and pseudoscalar nonets, respectively.
On the other hand, although the branching fraction 
${\cal B}_{\rm ex}(p\bar \kappa^0)\equiv{\cal B}(\Lambda_c^+ \to p \bar \kappa^0)
= (1.9 \pm 0.6) \times 10^{-3}$ is about an order of magnitude smaller than 
${\cal B}_{\rm ex}(\Lambda_c^+\to p\bar K^0)$~\cite{LHCb:2022sck,pdg}, 
its non-negligible size nevertheless supports an interpretation of 
these scalar channels within a short-distance framework.

The $SU(3)$ flavor $[SU(3)_f]$ symmetry has been extensively applied to heavy hadron decays~\cite{Chau:1995gk,Gronau:1995hm,Pan:2020qqo,Hsiao:2023mud,
Zhao:2018mov,Hsiao:2020iwc,Hsiao:2021nsc,Wang:2023uea,Kohara:1991ug,He:2018php,
He:2018joe,Wang:2020gmn,Cheng:2024lsn,Zeppenfeld:1980ex,Savage:1989qr,Savage:1991wu,
Sharma:1996sc,Lu:2016ogy,Geng:2017mxn,Geng:2017esc,
Geng:2018bow,Geng:2018plk,Wang:2017gxe,Wang:2019dls,
Jia:2019zxi,Huang:2021aqu,Wang:2022kwe,Xing:2023dni,Zhong:2022exp,He:2000ys,
Li:2012cfa,Cheng:2019ggx}.
A particularly advantageous formulation is
the $SU(3)_f$-based topological-diagram approach (TDA)~\cite{Chau:1995gk,Gronau:1995hm,
Pan:2020qqo,Hsiao:2023mud,Zhao:2018mov,Hsiao:2020iwc,Hsiao:2021nsc,
Wang:2023uea,Kohara:1991ug,He:2018php,
He:2018joe,Wang:2020gmn,Cheng:2024lsn,Li:2012cfa,Cheng:2019ggx}, which provides
an intuitive and diagrammatic framework to describe
short-distance $W$-boson emission and exchange topologies.
An equally useful formulation is the irreducible $SU(3)_f$ approach~(IRA)~\cite{Zeppenfeld:1980ex,Savage:1989qr,Savage:1991wu,
Sharma:1996sc,Lu:2016ogy,Geng:2017mxn,Geng:2017esc,
Geng:2018bow,Geng:2018plk,Wang:2017gxe,Wang:2019dls,
Jia:2019zxi,Huang:2021aqu,Wang:2022kwe,Xing:2023dni,Zhong:2022exp,He:2000ys},
which systematically organizes amplitudes according
to the irreducible group-theoretical representations.
Moreover, the TDA and IRA can be unified~\cite{He:2018php,He:2018joe,
Hsiao:2020iwc,Hsiao:2021nsc,Wang:2023uea,Cheng:2024lsn},
leading to a reduction in the number of independent parameters
and thereby facilitating numerical analyses of charm baryon decays.
We therefore propose extending the TDA framework to the study of ${\bf B}_c \to {\bf B} S$ processes.
This allows for a systematic examination of whether short-distance contributions,
represented by topological diagrams, dominate these decays.
At the same time, it provides a promising strategy to probe the internal structure
of light scalar mesons, helping to distinguish between
conventional $q\bar q$ $p$-wave states and exotic $q^2\bar q^2$ tetraquark configurations.

\section{Formalism}
The effective Hamiltonian that governs the weak transition
of the charm quark in the charmed baryon decays of ${\bf B}_c\to{\bf B}S$
is given by~\cite{Buchalla:1995vs,Buras:1998raa}
\begin{eqnarray}\label{Heff}
{\cal H}_{eff}&=&
\frac{G_F}{\sqrt 2}V_{uq'} V^*_{cq}[c_1(\bar u q' )(\bar q c)
+c_2 (\bar u_\beta q'_\alpha )(\bar q_\alpha c_\beta)]\,,
\end{eqnarray}
where $G_F$ is the Fermi constant, $c_{1,2}$ are the Wilson coefficients, and
the relevant Cabibbo-Kobayashi-Maskawa (CKM) matrix elements $V_{uq'} V^*_{cq}$ can be
$V_{ud} V^*_{cs}$, $V_{uq} V^*_{cq}$ ($q=d$ or $s$), and $V_{us} V^*_{cd}$.
Additionally, $(\bar q q')=\bar q\gamma_\mu(1-\gamma_5)q'$ denotes the quark current,
and the subscripts $(\alpha,\beta)$ are the color indices.
In Eq.~(\ref{Heff}), we neglect the penguin operators $V_{ub}V_{cb}^* c_j O_j$ ($j=3,4,...,6$).
This approximation is well justified by the strong CKM hierarchy~\cite{pdg}:
$|V_{ub}V_{cb}^*|\sim 10^{-4}$, which is much smaller than  
$|V_{us}V_{cd}^*|\sim 0.05$, $|V_{uq}V_{cq}^*|\sim 0.2$, and $|V_{ud}V_{cs}^*|\sim 1$. Moreover,
the penguin Wilson coefficients $c_j$ arise only at the loop level and
are therefore additionally suppressed by the QCD factor $\alpha_s/(4\pi)\sim 10^{-2}$.
As a result, for the purpose of analyzing branching fractions, 
it is an excellent approximation to retain only the tree-level operators,
which provide the dominant contributions.

To facilitate the analysis within the framework of $SU(3)_f$ symmetry,
we omit the Lorentz and color indices and express the Hamiltonian as
${\cal H}_{c}={\cal H}_{eff}/(G_F/\sqrt 2)=H_j^{ki}$,
where the indices $i$, $j$, and $k$, ranging from 1 to 3,
correspond to the light quark flavors $(u,d,s)$.
This tensor representation encodes the flavor transition $c\to q_i \bar q_j q_k$.
The non-zero components of $H_j^{ki}$
are~\cite{Pan:2020qqo,Hsiao:2020iwc,Hsiao:2021nsc,Wang:2023uea}
\begin{eqnarray}\label{Hijk}
H^{31}_2=V^*_{cs}V_{ud}\,,\quad
H^{31}_3=V^*_{cs}V_{us}\,,\quad
H^{21}_2=V^*_{cd}V_{ud}\,,\quad
H^{21}_3=V^*_{cd}V_{us}\,,\quad
\end{eqnarray}
corresponding to the charm quark decays
$c\to u \bar d s$, $u\bar s s$, $u \bar d d$, and $u \bar s d$, respectively.
The associated CKM matrix elements are approximately given by
$V^*_{cs}V_{ud}\simeq 1$, $V^*_{cs}V_{us}\simeq s_c$,
$V^*_{cd}V_{ud}\simeq -s_c$, and $V^*_{cd}V_{us}\simeq -s_c^2$,
where $s_c\equiv \sin\theta_c\simeq 0.22$, and $\theta_c$ is the Cabibbo angle.
In the irreducible forms of $SU(3)_f$, the charmed baryon anti-triplet ${\bf B}_c$
and the baryon octet ${\bf B}$ are given by
\begin{eqnarray}\label{3c8B}
{\bf B}_c=\left(\begin{array}{ccc}
0& \Lambda_c^+ & \Xi_c^+\\
-\Lambda_c^+&0&\Xi_c^0 \\
-\Xi_c^+&-\Xi_c^0&0
\end{array}\right)\,,\;
{\bf B}=\left(\begin{array}{ccc}
\frac{1}{\sqrt{6}}\Lambda^0+\frac{1}{\sqrt{2}}\Sigma^0 & \Sigma^+ & p\\
 \Sigma^- &\frac{1}{\sqrt{6}}\Lambda^0 -\frac{1}{\sqrt{2}}\Sigma^0  & n\\
 \Xi^- & \Xi^0 &-\sqrt{\frac{2}{3}}\Lambda^0
\end{array}\right)\,.
\end{eqnarray}
The light scalar final states, interpreted either as conventional $q\bar q$ p-wave mesons or
as exotic $q^2\bar q^2$ tetraquark states, can be organized into
two distinct irreducible representations under $SU(3)_f$ symmetry.
In the $q\bar q$ configuration, the light scalar mesons $S$
can be decomposed as~\cite{Stone:2013eaa,Maiani:2004uc}
\begin{eqnarray}\label{S_2q}
&&
f_0= \cos\theta_I|s\bar s\rangle+\sin\theta_I|\sqrt {1/2}(u\bar u+d\bar d)\rangle\,,\nonumber\\
&&
\sigma_0=-\sin\theta_I|s\bar s\rangle +\cos\theta_I |\sqrt {1/2}(u\bar u+d\bar d)\rangle\,,\nonumber\\
&&
a_0^+=|u\bar d\rangle\,,
a_0^0=|\sqrt {1/2}(u\bar u-d\bar d)\rangle\,,
a_0^-=|d\bar u\rangle\,,\nonumber\\
&&
\kappa^+=|u\bar s\rangle\,,
\kappa^0=|d\bar s\rangle\,,
\bar\kappa^0=|s\bar d\rangle\,,
\kappa^-=|s\bar u\rangle\,,
\end{eqnarray}
where  $\phi_I$ is the mixing angle in the $f_0-\sigma_0$ mixing system.
The corresponding $S$ nonet takes the form
\begin{eqnarray}\label{M_2q}
S_{q\bar q}=\left(\begin{array}{ccc}
\frac{1}{\sqrt{2}}(a_0^0+ c\phi_I\sigma_0 +s\phi_I f_0) & a_0^+ & \kappa^+\\
 a_0^- & \frac{-1}{\sqrt{2}}(a_0^0- c\phi_I\sigma_0 -s\phi_I f_0) & \kappa^0\\
 \kappa^- & \bar\kappa^0& -s\phi_I\sigma_0 +c\phi_I f_0
\end{array}\right)\,,
\end{eqnarray}
with $(c\phi_I,s\phi_I)=(\cos\phi_I,\sin\phi_i)$.
In the $q^2\bar q^2$ structure, the scalar mesons are instead described by~\cite{Maiani:2004uc}
\begin{eqnarray}\label{S_4q}
&&
f_0= \cos\theta_{II}|\sqrt {1/2}(u\bar u+d\bar d) s\bar s\rangle
+\sin\theta_{II}|u\bar u d\bar d\rangle\,,\;\nonumber\\
&&
\sigma_0=-\sin\theta_{II}|\sqrt {1/2}(u\bar u+d\bar d) s\bar s\rangle
+\cos\theta_{II} |u\bar u d\bar d\rangle\,,\;\nonumber\\
&&
a_0^+=|u\bar d s\bar s\rangle\,,\;
a_0^0=|\sqrt {1/2}(u\bar u-d\bar d)s\bar s\rangle\,,
a_0^-=|d\bar u s\bar s\rangle\,,\nonumber\\
&&
\kappa^+=|u\bar{s}d\bar{d}\rangle\,,\;
\kappa^0=|d\bar{s}u\bar{u}\rangle\,,\;
\bar\kappa^0=|s\bar{d}u\bar{u}\rangle\,,\;
\kappa^-=|s\bar{u}d\bar{d}\rangle\,,
\end{eqnarray}
where $\theta_{II}$ is the mixing angle in this scheme.
The corresponding light scalar nonet is then given by
\begin{eqnarray}\label{M_4q}
S_{q^2\bar q^2}=\left(\begin{array}{ccc}
\frac{1}{\sqrt{2}}(a_0^0+ c\phi_{II}f_0-s\phi_{II} \sigma_0) & a_0^+ & \kappa^+\\
 a_0^- & \frac{-1}{\sqrt{2}}(a_0^0-c\phi_{II}f_0+s\phi_{II} \sigma_0) & \kappa^0\\
 \kappa^- & \bar\kappa^0& s\phi_{II}f_0 +c\phi_{II} \sigma_0
\end{array}\right)\,,
\end{eqnarray}
with $(c\phi_{II},s\phi_{II})=(\cos\phi_{II},\sin\phi_{II})$.
%
\begin{figure}
\includegraphics[width=.38\textwidth]{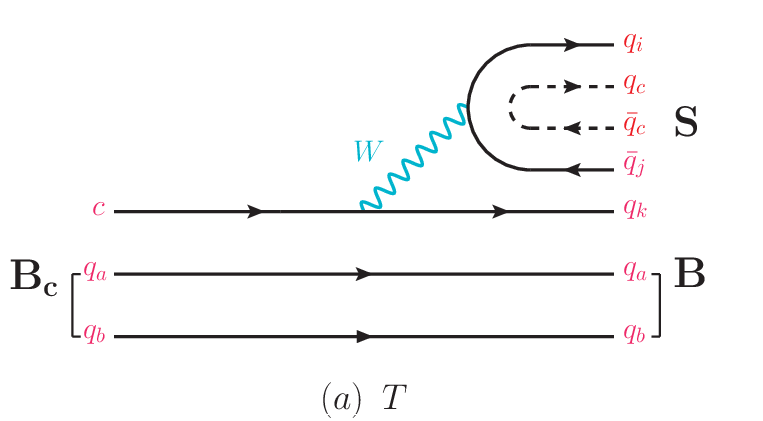}
\includegraphics[width=.38\textwidth]{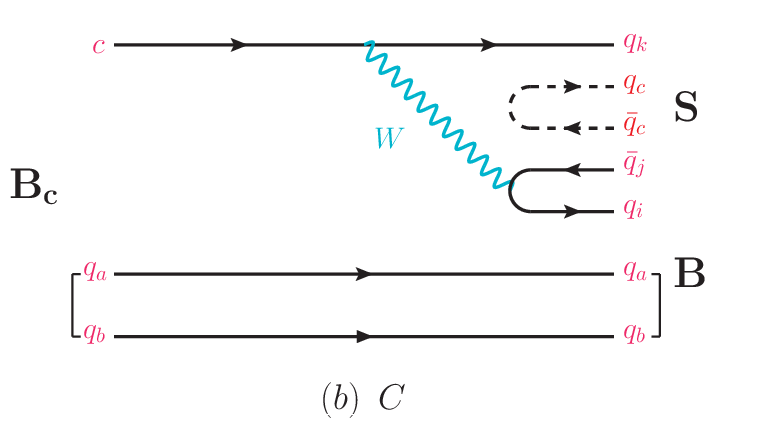}
\includegraphics[width=.38\textwidth]{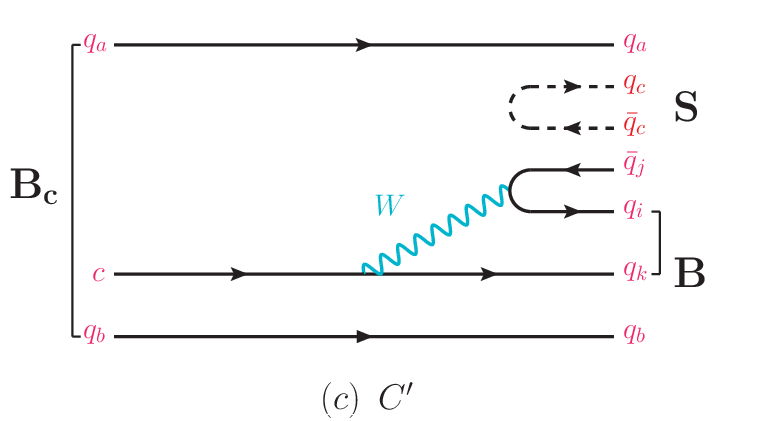}
\includegraphics[width=.38\textwidth]{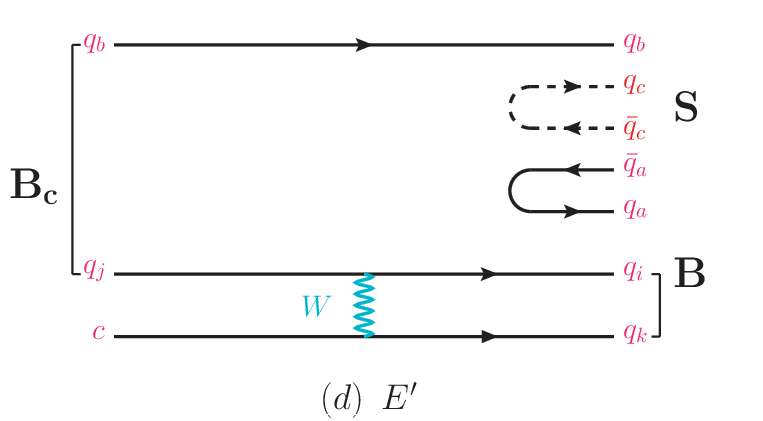}
\includegraphics[width=.38\textwidth]{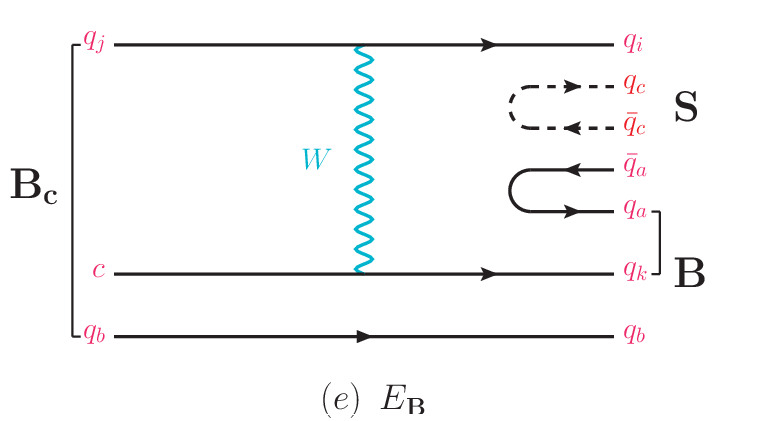}
\includegraphics[width=.38\textwidth]{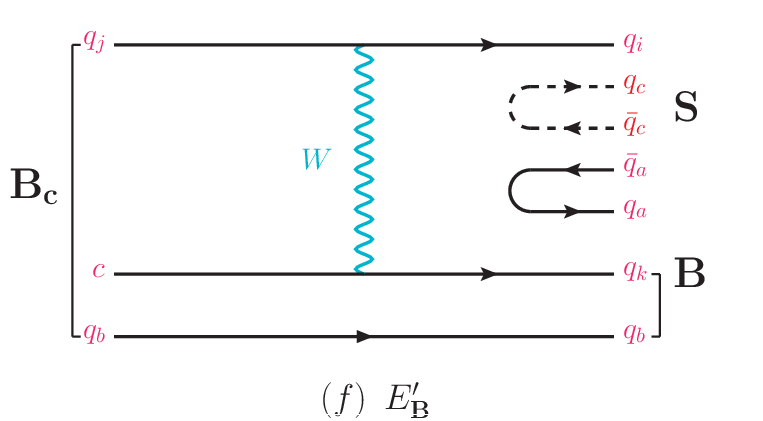}
\includegraphics[width=.38\textwidth]{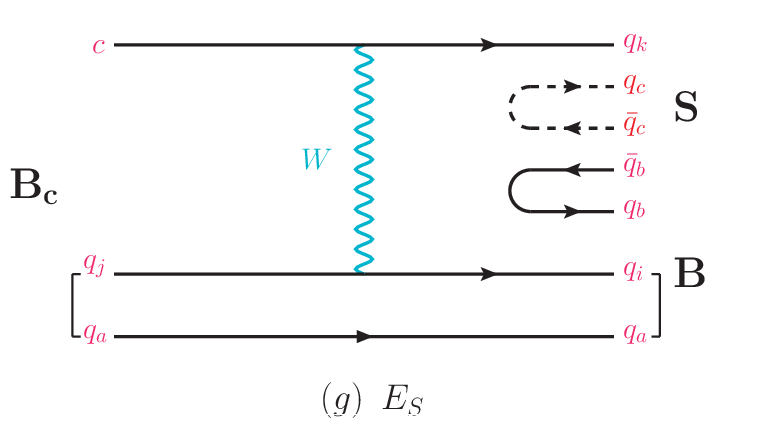}
\includegraphics[width=.38\textwidth]{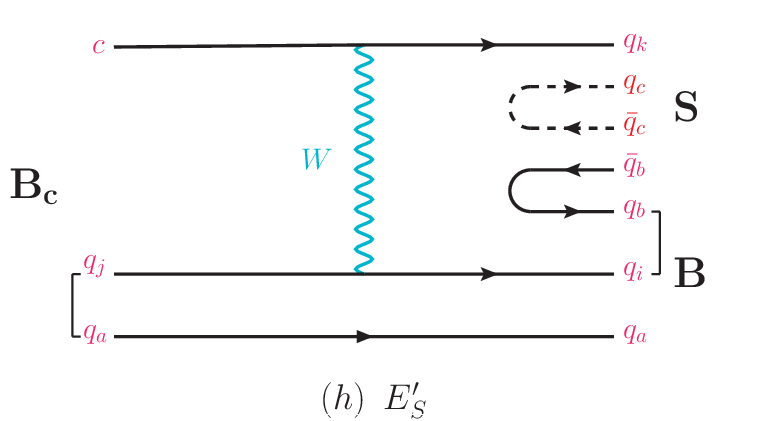}
\caption{
Topological diagrams for ${\bf B}_c \to {\bf B}S$ decays are presented,
featuring the possible inclusion of a $q_c \bar q_c$ pair
within the $q^2 \bar q^2$ configuration of the light scalar meson $S$.
Diagrams $(a)$, $(b)$, and $(c)$ represent $W$-emission processes,
while $(d)$ through $(h)$ correspond to $W$-exchange contributions.
The notations ``[''  and ``]''  indicate the anti-symmetric quark orderings
in the charmed baryon and octet baryon states, respectively.}\label{fig1}
 \end{figure}
%
The decays ${\bf B}_c\to{\bf B}S$ proceed through various topological diagrams,
as drawn in Fig.~\ref{fig1}. Specifically, Figs.~\ref{fig1}$(a)$ and Figs.~\ref{fig1}$(b)$
illustrate the external and internal $W$-boson emission processes, denoted as $T$ and $C$,
respectively. These topologies involve the ${\bf B}_c\to {\bf B}$ transition,
accompanied by the production of a light scalar meson from the vacuum.
Fig.~\ref{fig1}c shows the internal $W$-emission process labeled $C'$, which is non-factorizable.
In Fig.~\ref{fig1}$(d)$, the topology labeled $E'$ represents a $W$-boson exchange
involving only the quark lines of the final-state baryon,
without direct interaction with the quark contents of the light scalar meson.
The diagrams labeled $E_{\bf B}$ in Fig.~\ref{fig1}$(e)$ and $E_S$ in Fig.~\ref{fig1}$(g)$
illustrate $W$-exchange processes where the quark $q_k$, produced from the $c \to q_k$ transition,
is incorporated into ${\bf B}$ and $S$, respectively. The primed topologies,
$E'_{\bf B}$ and $E'_S$, share the same overall structure as $E_{\bf B}$ and $E_S$,
but differ in the specific anti-symmetric quark pair configuration within the baryon wave function.

Using the Hamiltonian components $H^{ki}_j$ from Eq.~(\ref{Hijk})
along with the irreducible forms of the initial and final states
in Eqs.~(\ref{3c8B}), (\ref{M_2q}), and (\ref{M_4q}),
we construct the decay amplitudes for ${\bf B}_c\to{\bf B}S$ processes.
This connection from the initial state ${\bf B}_c$ to the final-state baryon ${\bf B}$
and the light scalar meson $S$ leads to the TDA amplitudes,
expressed as~\cite{Kohara:1991ug,He:2018joe,Zhao:2018mov,
Pan:2020qqo,Hsiao:2020iwc,Wang:2023uea}
\begin{eqnarray}\label{amp1}
{\cal M}_{\rm TDA}({\bf B}_c\to{\bf B}S)
&=&
T{{\bf B}_c}^{ab} H^{ki}_j{\bf B}_{abk}S_j^i
+C {{\bf B}_c}^{ab} H^{ki}_j{\bf B}_{abi}S_j^k
+C^\prime{{\bf B}_c}^{ab} H^{ki}_j{\bf B}_{ikb}S_j^a\,\nonumber\\
&+&
E_{\bf B} {{\bf B}_c}^{jb} H^{ki}_j{\bf B}_{kab}S_a^i+
E_{\bf B}^{\prime}{{\bf B}_c}^{jb} H^{ki}_j{\bf B}_{kba}S_a^i\,\nonumber\\
&+&
E_{S}{{\bf B}_c}^{jb} H^{ki}_j{\bf B}_{iba}S_a^k+
E_{S}^{\prime} {{\bf B}_c}^{jb} H^{ki}_j{\bf B}_{iab}S_a^k+
E^{\prime}{{\bf B}_c}^{jb} H^{ki}_j{\bf B}_{ika}S_a^b\,.
\end{eqnarray}
These terms correspond to the topological diagrams in Figs.~\ref{fig1}(a)-(h), and are
parameterized by the amplitudes $(T, C^{(\prime)}, E_{\bf B}^{(\prime)}, E_{S}^{(\prime)}, E')$.
In deriving the decomposition of the baryon octet, the identity
${\bf B}_{ijk} = \epsilon_{ijl} {\bf B}^l_k$ has been employed.
Both types of the light scalar nonets, $S_{q\bar q}$ in Eq.~(\ref{M_2q}) 
and $S_{q^2\bar q^2}$ in Eq.~(\ref{M_4q}), are taken into account.

Since the combination $(E' - E_{\bf B}')$
appears universally in the amplitudes, indicating a redundancy in the parameterization,
we simplify the amplitude structure by setting $E_{\bf B}' = 0$.
In contrast with the topologies $T$, $C$, and $C'$,
the $W$-exchange contributions $E_i=(E_{\bf B},E_{S}^{(\prime)},E')$ require
a vacuum-induced gluon to produce an additional quark-antiquark pair,
denoted by $g\to q\bar q$, where $q\bar q$ can be $u\bar u$, $d\bar d$, or $s\bar s$.
The quark and antiquark are subsequently incorporated into the final-state baryon $\bf B$
and scalar meson $S$, respectively. Under exact $SU(3)_f$ symmetry,
the $W$-exchange amplitudes with $g\to s\bar s$ are indistinguishable 
from those with $g\to u\bar u$ or $d\bar d$. To clarify, 
we denote the former by $E_i^{s}$, while retaining $E_i$ for the latter cases.
When $SU(3)_f$ symmetry is broken, $E_i^{s}$ and $E_i$ are no longer equal, 
and their difference may be regarded as $SU(3)_f$-breaking effect.

Under these considerations, the topological amplitudes ${\cal M}_{\rm TDA}$
listed in Tables~\ref{tab1} and~\ref{tab2} can be fully expanded, with
all allowed decay channels expressed in terms of the corresponding topological amplitudes. 
In addition, each amplitude is accompanied by a prefactor of $s_c^0$, $s_c^1$, or $s_c^2$,
corresponding to Cabibbo-allowed (CA), singly Cabibbo-suppressed (SCS),
or doubly Cabibbo-suppressed (DCS) decays, respectively,
in the $\Lambda_c^+ \to {\bf B}S$ processes.

%
%
\begin{table}[t]
\caption{Topological amplitudes of $\Lambda_c^+\to {\bf B}a_0,{\bf B}\kappa$.}\label{tab1}
{
\scriptsize
\begin{tabular}{|l|l|}
\hline
Decay mode
&$\;\;\;\;\;\;\;\;\;\;\;\;\;\;\;\;$${\cal M}_{q\bar q}={\cal M}_{q^2\bar q^2}$
\\
\hline\hline
$\Lambda_c^+ \to \Lambda^0 a_0^+$
&$-\frac{1}{\sqrt{6}}(4T+C^\prime-E_{{\bf B}}-E^\prime)$
\\
$\Lambda_c^+ \to \Sigma^0 a_0^+$
&$\frac{1}{\sqrt{2}}(C^\prime +E_{{\bf B}}-E^\prime)$
\\
$\Lambda_c^+ \to \Sigma^+ a_0^0$
&$-\frac{1}{\sqrt 2}(C^\prime +E_{{\bf B}}-E^\prime)$
\\
$\Lambda_c^+ \to \Xi^0 \kappa^+$
&$E^{\prime (s)}$
\\
$\Lambda_c^+ \to p \bar \kappa^0$
&$2C-E_S^\prime$
\\
$\Lambda_c^+ \to \Lambda^0 \kappa^+$
&$- \frac{1}{\sqrt{6}}(4T+C^\prime-E_{{\bf B}}^{(s)}+E^{\prime (s)})s_c$
\\
$\Lambda_c^+ \to \Sigma^0 \kappa^+$
&$\frac{1}{\sqrt{2}}(C^\prime+E_{{\bf B}}^{(s)})s_c$
\\
$\Lambda_c^+ \to \Sigma^+ \kappa^0$
&$(-E_S^{\prime (s)}+C^\prime)s_c$
\\
$\Lambda_c^+ \to n a_0^+$
&$(-2T-C^\prime+E^\prime)s_c$
\\
$\Lambda_c^+ \to p a_0^0$
&$\frac{1}{\sqrt 2}(2C-C^\prime-E_{{\bf B}}-E_S^\prime+E^\prime)s_c $
\\
$\Lambda_c^+ \to p \kappa^0$
&$(C^\prime-2C)s_c^2$
\\
$\Lambda_c^+ \to n  \kappa^+$
&$-(C^\prime +2T)s_c^2$
\\
\hline
\end{tabular}}
\end{table}
%
%
\begin{table}[t]
\caption{Topological amplitudes of $\Lambda_c^+\to {\bf B}f_0,{\bf B}\sigma_0$
in the $f_0-\sigma_0$ mixing scheme.}\label{tab2}
{\scriptsize
\begin{tabular}{|l|l|l|}
\hline
Decay mode
&$\;\;\;\;\;\;\;\;\;\;\;\;\;\;\;\;\;\;\;\;\;\;\;\;$ ${\cal M}_{q\bar q}$
&$\;\;\;\;\;\;\;\;\;\;\;\;\;\;\;\;\;\;\;\;\;\;\;\;$ ${\cal M}_{q^2\bar q^2}$ \\
\hline\hline
$\Lambda_c^+ \to \Sigma^+ f_0$
&$\frac{1}{\sqrt{2}}(C^\prime-E_{{\bf B}}+E^\prime)s\phi_I+E_S^{\prime (s)} c\phi_I$
&$\frac{1}{\sqrt 2}(C^\prime-E_{{\bf B}}+E^\prime)c\phi_{II}+E_S^{\prime (s)} s\phi_{II}$
\\
$\Lambda_c^+ \to \Sigma^+ \sigma_0$
&$\frac{1}{\sqrt{2}}(C^\prime-E_{{\bf B}}+E^\prime)c\phi_I-E_S^{\prime (s)} s\phi_I$
&$-\frac{1}{\sqrt 2}(C^\prime-E_{{\bf B}}+E^\prime)s\phi_{II}+E_S^{\prime (s)} c\phi_{II}$
\\
$\Lambda_c^+ \to p f_0$
&$[-\frac{1}{\sqrt 2}(2C-C^\prime-E_S^\prime+E_{\bf B}-E^\prime)s\phi_I+2Cc\phi_I]s_c$
&$[-\frac{1}{\sqrt 2}(2C-C^\prime-E_S^\prime+E_{\bf B}-E^\prime)c\phi_{II}+2Cs\phi_{II}]s_c$
\\
$\Lambda_c^+ \to p \sigma_0$
&$[-\frac{1}{\sqrt 2}(2C-C^\prime-E_S^\prime+E_{\bf B}-E^\prime)c\phi_I-2Cs\phi_I]s_c$
&$[\frac{1}{\sqrt 2}(2C-C^\prime-E_S^\prime+E_{\bf B}-E^\prime)s\phi_{II}+2Cc\phi_{II}]s_c$
\\
\hline
\end{tabular}}
\end{table}
%

Although the topological diagrammatic approach (TDA) 
offers an intuitive classification of all possible subprocesses, 
it does not constitute an irreducible $SU(3)_f$ framework. 
By contrast, the irreducible representation approach (IRA)
expresses the amplitude ${\cal M}_{\rm IRA}({\bf B}_c \to {\bf B}S)$ in terms of
seven independent parameters $\hat a_i$ $(i=1,2,\dots,7)$.
This formulation is based on decomposing the effective Hamiltonian in Eq.~(\ref{Heff}) 
into irreducible $SU(3)_f$ components, 
${\cal H}_{\text{IRA}}=(c_-/2)\epsilon^{ijl}H(6)_{lk}+c_+H(\overline{15})_{k}^{ij}$, 
with $c_\mp=(c_1\mp c_2)$. The non-zero entries of $H(6)_{lk}$ and $H(\overline{15})_k^{ij}$ 
are given in Refs.~\cite{Savage:1989qr,Hsiao:2020iwc}. As a consequence of this decomposition, 
the decay amplitude can be written as
${\cal M}_{\text{IRA}}({\bf B}_c \to {\bf B}S)={\cal M}_{6}+{\cal M}_{\overline{15}}$, 
where
\begin{eqnarray}
{\cal M}_{6}&=&
\hat a_1 H_{ij}(6)T^{ik}{\bf B}_k^l S_l^j+
\hat a_2 H_{ij}(6)T^{ik}S_k^l {\bf B}_l^j+
\hat a_3 H_{ij}(6){\bf B}_k^i S_l^j T^{kl}\,,\nonumber\\
{\cal M}_{\overline{15}}&=&
\hat a_4H_{k}^{li}(\overline{15}){\bf B}_{c\,j} S_i^j {\bf B}^l_k+
[\hat a_5{\bf B}^i_j S^l_i H(\overline{15})^{jk}_l
+\hat a_6{\bf B}^k_l S^i_j H(\overline{15})^{jl}_i 
+\hat a_7{\bf B}^l_i S^i_j H(\overline{15})^{jk}_l]{\bf B}_{c\,k}\,.
\end{eqnarray}
Here $T^{ij} \equiv{\bf B}_{c\,k}\epsilon^{ijk}$, and
$a_i$ ($i=1,2, ...,7$) denote the $SU(3)_f$ invariant amplitudes.
Owing to the irreducible decomposition of the effective Hamiltonian, the IRA formulation
contains no redundant parameters.
Since the equivalence between the TDA and IRA descriptions has been firmly established
in Refs.~\cite{He:2018php,He:2018joe,Hsiao:2020iwc,Hsiao:2021nsc,Wang:2023uea,Cheng:2024lsn},
one may exploit the relation
${\cal M}_{\rm TDA}({\bf B}_c\to {\bf B}S) = {\cal M}_{\rm IRA}({\bf B}_c\to {\bf B}S)$
to further reduce the number of independent topological amplitudes. 
In particular, this matching leads to the identification $E_S' = E_{\bf B} = \hat a_4$.
An analogous reduction also appears in ${\bf B}_c\to {\bf B}M$, where
$E_M^\prime = E_{\bf B}=a_4$~\cite{Hsiao:2021nsc}. 

The TDA amplitudes $T$ and $C$ correspond to
factorizable external and internal $W$-boson emission processes, respectively.
To estimate $T$ and $C$, we consider the factorizable amplitudes
${\cal M}_{\rm fac}(\Lambda_c^+\to\Lambda^0 \kappa^+)=a_1 f_{\kappa}q^\mu \bar u_\Lambda
(f_1^{\Lambda_c \Lambda}\gamma_\mu+g_1^{\Lambda_c \Lambda}\gamma_\mu\gamma_5)
u_{\Lambda_c}$ and ${\cal M}_{\rm fac}(\Lambda_c^+\to p \bar \kappa^0)=
a_2 f_{\kappa}q^\mu \bar u_p
(f_1^{\Lambda_c p}\gamma_\mu+g_1^{\Lambda_c p}\gamma_\mu\gamma_5)
u_{\Lambda_c}$, where $(a_1,a_2)=(0.9,0.4)$~\cite{Yu:2020vlt},
$|f_\kappa|=45.5\times 10^{-3}$~GeV~\cite{Cheng:2002ai,Cheng:2022vbw},
$(f_1^{\Lambda_c \Lambda},g_1^{\Lambda_c \Lambda})=
-\sqrt{2/3}(f_1^{\Lambda_c p},g_1^{\Lambda_c p})=(0.38,0.34)$~\cite{Hsiao:2019wyd}.
From this, we obtain
$|T|=\sqrt {3/8}|{\cal M}(\Lambda_c^+\to\Lambda^0 \kappa^+)|=0.045$~GeV$^3$ and
$|C|=|{\cal M}_{\rm fac}(\Lambda_c^+\to p \bar \kappa^0)|/2=0.022$~GeV$^3$.
Although $|f_\kappa|$ is the largest among the light scalar decay constants $f_S$,
the resulting factorizable contributions are still numerically negligible,
implying $T\simeq C\simeq 0$.

While all currently available data arise from $\Lambda_c^+$ decays,
we therefore restrict our analysis to the $\Lambda_c^+\to {\bf B}S$ decay channels.
With the simplifying assumptions $T=C=E_{\bf B}^\prime=0$ 
and the equivalence relation $E_S^\prime=E_{\bf B}$,
the set of topological amplitudes relevant to $\Lambda_c^+ \to {\bf B} S$ decays 
is reduced to
\begin{eqnarray}\label{remaining}
|C'|\,,\; |E'|e^{\delta_{E'}}\,,\;|E_{\bf B}|e^{i\delta_{E_{\bf B}}}\,,
\end{eqnarray}
where $\delta_{E'}$ and $\delta_{E_{\bf B}}$ denote the relative strong phases.
This parameterization involves a total of five independent parameters. The amplitude $E_S$,
which contributes exclusively to $\Xi_c^0 \to {\bf B} S$ decays, is not considered in the present analysis.

The three currently available data points
are insufficient to constrain the remaining parameters in Eq.~(\ref{remaining})
through the global fit. Nevertheless, the measured resonant branching fractions 
${\cal B}({\bf B}_c \to {\bf B}S, S \to MM)$ contain information on 
the underlying two-body branching fractions ${\cal B}({\bf B}_c \to {\bf B}S)$ 
and, in principle, allow their extraction. In particular, the total branching fractions
${\cal B}_{\rm T}^1(\Lambda_c^+ \to \Sigma^+ \pi^+\pi^-)=(44.7\pm 2.2)\times 10^{-3}$ and
${\cal B}_{\rm T}^2(\Lambda_c^+ \to\Sigma^+ K^+K^-)=(35.9\pm 3.5)\times 10^{-4}$~\cite{pdg}
are significantly larger than the corresponding non-resonant components,
${\cal B}_{\rm NR}^1(\Lambda_c^+ \to \Sigma^+ \pi^+\pi^-)=(28.0\pm 5.0)\times10^{-3}$ and
${\cal B}_{\rm NR}^2(\Lambda_c^+ \to \Sigma^+ K^+ K^-)
=(4.6\pm 0.4)\times10^{-4}$~\cite{Geng:2018upx,Geng:2024sgq}.
This clearly points to the contributions from
${\cal B}_{f_0}^{1(2)}\equiv {\cal B}[\Lambda_c^+ \to \Sigma^+ f_0,f_0\to \pi^+ \pi^-(K^+ K^-)]$,
${\cal B}_{\sigma_0}^1\equiv {\cal B}(\Lambda_c^+ \to \Sigma^+ \sigma_0,\sigma_0\to\pi^+\pi^-)$, and
${\cal B}_{a_0}^2\equiv {\cal B}(\Lambda_c^+ \to \Sigma^+ a_0^0, a_0^0\to K^+ K^-)$.
Additionally, the observed branching fractions
${\cal B}_T^3\equiv{\cal B}(\Lambda_c^+\to\Sigma^0 \pi^+\eta)
=(7.6\pm 0.8)\times 10^{-3}$~\cite{pdg} and
${\cal B}_T^4\equiv{\cal B}(\Lambda_c^+\to\Xi^0 K^0\pi^+)
=(6.1\pm 0.6)\times 10^{-3}$~\cite{BESIII:2025rda} point to the contributions from
${\cal B}_{a_0}^3\equiv{\cal B}(\Lambda_c^+\to\Sigma^0 a_0^+, a_0^+\to \pi^+\eta)$ and
${\cal B}_{\kappa}^4\equiv{\cal B}(\Lambda_c^+\to\Xi^0\kappa^+,\kappa^+\to K^0\pi^+)$.
Since the resonant modes
${\cal B}_{\rho^0}^1\equiv {\cal B}(\Lambda_c^+ \to \Sigma^+ \rho^0,\rho^0\to \pi^+\pi^-)$,
${\cal B}_{\omega}^1\equiv {\cal B}(\Lambda_c^+ \to \Sigma^+ \omega,\omega\to \pi^+\pi^-)$,
${\cal B}_{\phi}^2\equiv {\cal B}(\Lambda_c^+ \to \Sigma^+ \phi,\phi\to K^+ K^-)$, and
${\cal B}_{\Sigma^*}^1\equiv{\cal B}(\Lambda_c^+ \to \Sigma^{*0} \pi^+,\Sigma^{*0}\to\Sigma^+\pi^-)$
also contribute to ${\cal B}_{\rm T}^1$ and ${\cal B}_{\rm T}^2$, as well as
${\cal B}_{\Sigma^*}^3\equiv {\cal B}(\Lambda_c^+ \to \Sigma^{*+}\eta,\Sigma^{*+}\to \Sigma^0\pi^+)$
to ${\cal B}_{\rm T}^3$ and
${\cal B}_{K^*}^4\equiv {\cal B}(\Lambda_c^+ \to \Xi^0 K^{*+},K^{*+}\to K^0\pi^+)$ to ${\cal B}_{\rm T}^4$,
their contributions must be carefully estimated and subtracted from the totals.

The amplitudes
${\cal M}(\Lambda_c^+ \to \Sigma^+ \rho^0)$,
${\cal M}(\Lambda_c^+ \to \Sigma^+ \omega)$,
${\cal M}(\Lambda_c^+ \to\Sigma^+ \phi)$,
${\cal M}(\Lambda_c^+ \to \Sigma^{*0} \pi^+)$,
${\cal M}(\Lambda_c^+ \to \Sigma^{*+} \eta)$, and
${\cal M}(\Lambda_c^+ \to \Xi^0 K^{*+})$ have been investigated
within the $SU(3)_f$ framework~\cite{Hsiao:2020iwc,Hsiao:2019yur}.
The updated analyses yield the branching fractions
$(4.5\pm 2.3)\times 10^{-3}$,
$(16.9\pm 4.3)\times 10^{-3}$,
$(4.0\pm 0.6)\times 10^{-3}$,
$(6.1\pm 0.5)\times 10^{-3}$,
$(7.8\pm 1.6)\times 10^{-3}$, and
$(6.2\pm 2.4)\times 10^{-3}$, respectively.
Within the same framework, the resonant amplitude is written as
${\cal M}(\Lambda_c^+ \to \Sigma^+ \rho^0,\rho^0\to\pi^+\pi^-)
={\cal M}(\rho^0\to \pi^+\pi^-)D_{\rho^0}^{-1}
{\cal M}(\Lambda_c^+ \to \Sigma^+ \rho^0)$~\cite{Wang:2025mdn},
where the Breit-Wigner (BW) propagator is given by
$D_{\rho^0}^{-1}=1/(q^2-m_\rho^2-im_\rho\Gamma_\rho)$, and
${\cal M}(\rho^0\to \pi^+\pi^-)$ describes the subsequent strong decay.
Analogous expressions apply to the resonant channels
${\cal M}(\Lambda_c^+ \to \Sigma^+ \omega,\omega\to\pi^+\pi^-)$,
${\cal M}(\Lambda_c^+ \to\Sigma^+ \phi,\phi\to K^+ K^-)$,
${\cal M}(\Lambda_c^+ \to \Sigma^{*0} \pi^+, \Sigma^{*0}\to \Sigma^+\pi^-)$,
${\cal M}(\Lambda_c^+ \to \Sigma^{*+} \eta, \Sigma^{*+}\to \Sigma^0\pi^+)$, and
${\cal M}(\Lambda_c^+ \to \Xi^0 K^{*+},K^{*+}\to K^0\pi^+)$.

In the limit $m_h\Gamma_h\to 0$, where $h$ denotes
$\rho^0$, $\omega$, $\phi$, $K^{*+}$, or $\Sigma^*$,
the squared propagator can be simplified
using the narrow-width approximation~(NWA)~\cite{Cheng:2010vk},
\begin{eqnarray}\label{NWA}
\frac{1}{(q^2-m_h^2)^2+m_h^2 \Gamma_h^2}
\simeq \frac{\pi}{m_h\Gamma_h}\delta(q^2-m_h^2)\,.
\end{eqnarray}
Applying this relation to the decay width of a three-body decay,
$d\Gamma=1/[(2\pi)^3 32 M_{\Lambda_c}^3]$
$|{\cal M}(\Lambda_c^+\to{\bf B}(p_1)M(p_2)M(p_3)|^2 dm_{12}^2 dm_{23}^2$~\cite{pdg},
with $m_{12}^2=(p_1+p_2)^2$ and $m^2_{23}=(p_2+p_3)^2$,
the NWA simplifies the resonant three-body branching fractions as
\begin{eqnarray}\label{resBxB}
{\cal B}_V&\equiv&{\cal B}({\bf B}_c\to {\bf B}V,V\to MM)\simeq
{\cal B}({\bf B}_c\to {\bf B}V)\times {\cal B}(V\to MM)\,,\nonumber\\
{\cal B}_{{\bf B}^*}&\equiv&{\cal B}({\bf B}_c\to {\bf B}^* M, {\bf B}^*\to {\bf B}M)\simeq
{\cal B}({\bf B}_c\to {\bf B}^* M)\times  {\cal B}({\bf B}^*\to {\bf B}M)\,.
\end{eqnarray}
Using $SU(3)_f$ amplitudes and the known decay fractions of the resonances, we obtain
${\cal B}_{\rho^0}^1\simeq{\cal B}(\Lambda_c^+ \to \Sigma^+ \rho^0)\times{\cal B}(\rho^0\to \pi^+\pi^-)
=(4.5\pm 2.3)\times 10^{-3}$,
${\cal B}_{\omega}^1\simeq{\cal B}(\Lambda_c^+ \to \Sigma^+ \omega)\times{\cal B}(\omega\to \pi^+\pi^-)
=(0.26\pm 0.07)\times 10^{-3}$,
${\cal B}_{\phi}^2\simeq{\cal B}(\Lambda_c^+ \to \Sigma^+ \phi)\times {\cal B}(\phi\to K^+ K^-)
=(2.0\pm 0.3)\times 10^{-3}$,
${\cal B}_{\Sigma^*}^1\simeq{\cal B}(\Lambda_c^+ \to \Sigma^{*0} \pi^+)
\times {\cal B}(\Sigma^{*0}\to\Sigma^+\pi^-)=(3.6\pm 0.5)\times 10^{-4}$,
${\cal B}_{\Sigma^*}^3\simeq{\cal B}(\Lambda_c^+ \to \Sigma^{*+} \eta)
\times {\cal B}(\Sigma^{*+}\to\Sigma^0\pi^+)=(3.6\pm 0.5)\times 10^{-4}$,
and
${\cal B}_{K^*}^4\simeq{\cal B}(\Lambda_c^+ \to \Xi^0 K^{*+})
\times {\cal B}(K^{*+}\to K^0\pi^+)=(4.1\pm 1.6)\times 10^{-3}$.
Here, we have used~\cite{pdg}:
${\cal B}(\rho^0\to \pi^+\pi^-)\simeq 1$,
${\cal B}(\omega\to \pi^+\pi^-)=(1.53\pm 0.13)\times 10^{-2}$,
${\cal B}(\phi\to K^+ K^-)=(49.1\pm 0.5)\times 10^{-2}$,
${\cal B}(\Sigma^{*0(+)}\to\Sigma^{+(0)}\pi^{-(+)})=(5.85\pm 0.75)\times 10^{-2}$, and
${\cal B}(K^{*+}\to K^0 \pi^+)=0.67$.

In practice, the quantity $m_h\Gamma_h$ is not always negligibly small; for example,
$m_\rho \Gamma_\rho\simeq 0.12$~GeV$^2$. By comparing both sides of Eq.~(\ref{NWA}),
we find that the NWA underestimates the full BW integral
by 13\% for ${\cal B}_{\rho^0}^1$, while it overestimates ${\cal B}_{\omega}^1$ by 0.7\%,
${\cal B}_\phi^2$ by 2\%, ${\cal B}_{\Sigma^*}^{1,3}$ by 13\%, and ${\cal B}_{K^*}^4$ by 14\%.
The particularly high accuracy for ${\cal B}_{\omega}^1$
can be traced to the small width of the $\omega$ resonance,
$m_\omega \Gamma_\omega\simeq 0.007$~GeV$^2$.
Using the relations
${\cal B}_{\rm T}^1={\cal B}_{\rm NR}^1+{\cal B}_{f_0}^1+{\cal B}_{\sigma_0}^1
+{\cal B}_{\rho^0}^1+{\cal B}_{\omega}^1+{\cal B}_{\Sigma^*}^1$,
${\cal B}_{\rm T}^2={\cal B}_{\rm NR}^2+{\cal B}_{f_0}^2+{\cal B}_{a_0^0}^2+{\cal B}_{\phi}^2$,
${\cal B}_{\rm T}^3={\cal B}_{a_0}^3+{\cal B}_{\Sigma^*}^3$, and
${\cal B}_{\rm T}^4={\cal B}_{\kappa}^4+{\cal B}_{K^*}^4$,
we extract the scalar meson resonance contributions as
${\cal B}_{\rm ex(res)}(f_0,\sigma_0)\equiv{\cal B}_{f_0}^1+{\cal B}_{\sigma_0}^1=(11.0\pm6.0)\times10^{-3}$,
${\cal B}_{\rm ex(res)}(f_0,a_0)\equiv{\cal B}_{f_0}^2+{\cal B}_{a_0}^2=(10.8\pm 4.6)\times10^{-4}$,
${\cal B}_{\rm ex(res)}(a_0^+)\equiv{\cal B}_{a_0}^3=(7.1\pm 0.7)\times10^{-3}$, and
${\cal B}_{\rm ex(res)}(\kappa^+)\equiv{\cal B}_{\kappa}^4=(3.9\pm 1.9)\times10^{-3}$,
where the corrections to the NWA values have been incorporated.

\section{Numerical analysis}
In our numerical analysis, we take
$s_c=\lambda=0.22453\pm 0.00044$ from the PDG~\cite{pdg},
where $\lambda$ denotes the Wolfenstein parameter.
For the $f_0-\sigma_0$ mixing, two scenarios are considered.
In the $q\bar{q}$ configuration, the mixing angle $\theta_I$
is adopted as a weighted average of the values reported in various studies~\cite{Hsiao:2023qtk,Ochs:2013gi,Li:2012sw,LHCb:2015klp,
Anisovich:2002wy,DiDonato:2011kr,Ochs:2013gi,Minkowski:1998mf,
Sarantsev:2021ein,Klempt:2021nuf,Braghin:2022uih}.
In the alternative $q^2\bar{q}^2$ configuration,
the mixing angle $\theta_{II}$ is taken from Ref.~\cite{Maiani:2004uc}.
Numerically, we obtain
\begin{eqnarray}\label{theta12}
\theta_{I}&=&(156.7\pm 0.7)^\circ\,,\;
\nonumber\\
\theta_{II}&=&(174.6^{+3.4}_{-3.2})^\circ\,.
\end{eqnarray}
The experimental inputs used in the analysis are summarized as
\begin{eqnarray}\label{6data}
&&
{\cal B}_{\rm ex}(\Lambda a_0^+)\equiv{\cal B}(\Lambda_c^+ \to \Lambda a_0^+)
=(1.23\pm 0.21)\times 10^{-2}\,,\nonumber\\
&&
{\cal B}_{\rm ex}(p f_0)\equiv{\cal B}(\Lambda_c^+ \to p f_0)= (3.5 \pm 2.3) \times 10^{-3}\,,
\nonumber\\
&&
{\cal B}_{\rm ex}(p\bar \kappa^0)\equiv{\cal B}(\Lambda_c^+ \to p \bar \kappa^0)
= (1.9 \pm 0.6) \times 10^{-3}\,,
\nonumber\\
&&
{\cal B}_{\rm ex(res)}(f_0,\sigma_0)\equiv{\cal B}_{f_0}^1+{\cal B}_{\sigma_0}^1
=(11.0\pm6.0)\times10^{-3}\,,\nonumber\\
&&
{\cal B}_{\rm ex(res)}(f_0,a_0)\equiv{\cal B}_{f_0}^2+{\cal B}_{a_0}^2
=(10.8\pm 4.6)\times10^{-4}\,,\nonumber\\
&&
{\cal B}_{\rm ex(res)}(a_0^+)\equiv{\cal B}_{a_0}^3
=(7.1\pm 0.7)\times10^{-3}\,,\nonumber\\
&&
{\cal B}_{\rm ex(res)}(\kappa^+)\equiv{\cal B}_{\kappa}^4
=(3.9\pm 1.9)\times10^{-3}\,.
\end{eqnarray}
In Eq.~(\ref{6data}), the quantities
${\cal B}_{\rm ex(res)}(f_0,a_0)$, ${\cal B}_{\rm ex(res)}(f_0,\sigma_0)$, ${\cal B}_{\rm ex(res)}(a_0^+)$ and
${\cal B}_{\rm ex(res)}(\kappa^+)$ are extracted from three-body resonant branching fractions
or their combinations. These observables can be related to the corresponding two-body branching fractions 
by properly accounting for the resonance effects. 
To establish this connection, we start with the resonant amplitude,
which can be written as~\cite{Wang:2025mdn}
\begin{eqnarray}\label{3bAmp}
{\cal M}(\Lambda_c^+ \to {\bf B} S,S\to MM)
={\cal M}(S\to MM)D_S^{-1}{\cal M}(\Lambda_c^+ \to {\bf B} S)\,,
\end{eqnarray}
where $D_S^{-1}$ denotes the resonance propagator.
For the decays $f_0\to(\pi\pi,K\bar K)$ and $a_0\to(\eta\pi, K\bar K)$,
the presence of multiple coupled channels, together with
the broad decay widths $\Gamma_{\sigma_0}\sim m_{\sigma_0}$ and
$\Gamma_{\kappa}\sim m_{\kappa}$,
renders the use of the conventional Breit–Wigner propagator $D_h^{-1}$
and the narrow-width approximation invalid.

We employ the Flatte parameterization for the $f_0$ and $a_0$ resonances,
and an energy-dependent-width Breit-Wigner form for the broad $\sigma_0$ and $\kappa$ states~\cite{Flatte:1976xu,Flatte:1976xv,Baru:2004xg,pdg,Ceci:2013zta}
\begin{eqnarray}\label{RS0}
&&
D_{f_0}=m_{f_0}^2-q^2-im_{f_0}(g_{f_0 \pi\pi}\rho_{\pi\pi}+g_{f_0 KK}\rho_{KK}F_{KK}^2)\,,\nonumber\\
&&
D_{a_0}=m_{a_0}^2-q^2-i(g_{a_0\pi\eta}^2\rho_{\pi\eta}+g_{a_0 KK}^2\rho_{KK})\,,\nonumber\\
&&
D_{\sigma_0}=
{m_{\sigma_0}^2-q^2-\Gamma_{\sigma_0}^2(q^2)/4-i\,m_{\sigma_0}\Gamma_{\sigma_0}(q^2)}\,,
\nonumber\\
&&
D_{\kappa}={m_{\kappa}^2-q^2-\Gamma_{\kappa}^2(q^2)/4-i\,m_{\kappa}\Gamma_{\kappa}(q^2)}\,,
\end{eqnarray}
where $q^2\equiv (p_{M}+p_{M})^2$, $\rho_{M M}=[(1-m_+^2/q^2)(1-m_-^2/q^2)]^{1/2}$
with $m_{\pm}=m_{M}\pm m_{M}$, and $\Gamma_{S}(q^2)
\equiv (\rho_{MM}/\bar\rho_{MM})\Gamma^0_{S}$, with $\bar\rho_{MM}$ evaluated at $q^2=m_{S}^2$.
To regulate the kaon-loop phase space, we include a phenomenological form factor
$F_{KK}=e^{-\alpha k^2}$ with $\alpha=2$~GeV$^{-2}$~\cite{LHCb:2014ooi,LHCb:2014vbo},
where $k$ denotes the kaon momentum in the $KK$ rest frame.
This form factor suppresses the unphysical growth of $\rho_{KK}$ far above the $KK$ threshold~\cite{Bugg:2008ig,LHCb:2014ooi,LHCb:2014vbo}.
The resonance parameters entering Eq.~(\ref{RS0})
are summarized as~\cite{LHCb:2014ooi,LHCb:2014vbo,Bugg:2008ig,
Abele:1998qd,BaBar:2005vhe,Sarantsev:2021ein,LHCb:2019sus,BESIII:2016tdb}
\begin{eqnarray}\label{coupling1}
&&
(m_{f_0},g_{f_0 \pi\pi})=(949.9\pm 2.1,167\pm 8)~\text{MeV}\,,\;
g_{f_0 KK}=(3.05\pm 0.13)g_{f_0 \pi\pi}\,,\nonumber\\
&&
(m_{a_0},g_{a_0 \pi\eta})=(999\pm 2,324\pm 15)~\text{MeV}\,,\;
g^2_{a_0 KK}=(1.03\pm 0.14)g^2_{a_0 \pi\eta}\,,\nonumber\\
&&
(m_{\sigma_0},\Gamma_{\sigma_0}^0)=
(527.0\pm 7.7,513.5\pm 15.2)~\text{MeV}\,,\nonumber\\
&&
(m_{\kappa},\Gamma_{\kappa}^0)=
(838\pm 11,463\pm 27)~\text{MeV}\,.
\end{eqnarray}

In Eq.~(\ref{3bAmp}), the decay amplitude ${\cal M}(S\to MM)=C_{SMM}$
is characterized by the strong coupling constant $C_{SMM}$.
The explicit expressions of these couplings are given by~\cite{Baru:2004xg,Bugg:2008ig}
\begin{eqnarray}\label{coupling2}
&&
C_{f_0\pi\pi}=(8\pi m_{f_0} g_{f_0 \pi\pi})^{1/2}\,,\;
C_{f_0 KK}=(8\pi m_{f_0} g_{f_0 KK})^{1/2}\,,\nonumber\\
&&
C_{a_0\pi \eta}=4\sqrt{\pi}g_{a_0 \pi\eta}\,,\;
C_{a_0 KK}=4\sqrt{\pi}g_{a_0 KK}\,,\nonumber\\
&&
C_{\sigma_0\pi\pi}=
[(8\pi m_{\sigma_0}^2/|\vec{p}_{\rm cm}|)(2/3)\Gamma_{\sigma_0}^0]^{1/2}\,,\;\nonumber\\
&&
C_{\kappa K\pi}=
[(8\pi m_{\kappa}^2/|\vec{p}_{\rm cm}|)(2/3)\Gamma_{\kappa}^0]^{1/2}\,.
\end{eqnarray}

Under the $SU(3)_f$ framework,
${\cal B}_{\rm th(res)}(f_0,\sigma_0)$, ${\cal B}_{\rm th(res)}(f_0,a_0)$, 
${\cal B}_{\rm th(res)}(a_0^+)$, and
${\cal B}_{\rm th(res)}(\kappa^+)$ can be expressed as
\begin{eqnarray}\label{3bB}
&&
{\cal B}_{\rm th(res)}(f_0,\sigma_0)=
\bar\alpha_1 {\cal B}_{\rm th}(\Lambda_c^+ \to \Sigma^+ f_0)+
\bar\alpha_2{\cal B}_{\rm th}(\Lambda_c^+ \to \Sigma^+ \sigma_0)\,,
\nonumber\\
&&
{\cal B}_{\rm th(res)}(f_0,a_0)=
\bar\alpha_3{\cal B}_{\rm th}(\Lambda_c^+ \to \Sigma^+ f_0)+
\bar\alpha_4{\cal B}_{\rm th}(\Lambda_c^+ \to \Sigma^+ a_0^0)\,,
\nonumber\\
&&
{\cal B}_{\rm th(res)}(a_0^+)=
\bar\alpha_5{\cal B}_{\rm th}(\Lambda_c^+ \to \Sigma^0 a_0^+)\,,\;
\nonumber\\&&
{\cal B}_{\rm th(res)}(\kappa^+)=
\bar\alpha_6{\cal B}_{\rm th}(\Lambda_c^+ \to \Xi^0 \kappa^+)\,.
\end{eqnarray}
In Eq.~(\ref{3bB}), $(\bar\alpha_1,\bar\alpha_2)$ 
are introduced as theoretical inputs to account for the resonant decays of $(f_0,\sigma_0)\to\pi^+\pi^-$.
Accordingly, the branching fractions satisfy
${\cal B}_{\rm th}(\Lambda_c^+ \to \Sigma^+ f_0,f_0\to\pi^+\pi^-)
=\bar\alpha_1{\cal B}_{\rm th}(\Lambda_c^+ \to \Sigma^+ f_0)$ and
${\cal B}_{\rm th}(\Lambda_c^+ \to \Sigma^+ \sigma_0,\sigma_0\to\pi^+\pi^-)
=\bar\alpha_2{\cal B}_{\rm th}(\Lambda_c^+ \to \Sigma^+ \sigma_0)$. Similarly,
$(\bar\alpha_3,\bar\alpha_4)$ correspond to the resonant transitions $(f_0,a_0^0)\to K^+ K^-$,
$\bar\alpha_5$ describes $a_0^+\to \pi^+\eta$, and
$\bar\alpha_6$ accounts for $\kappa^+\to K^0\pi^+$. 
The coefficients $\alpha_j$ ($j=1,2,...,6$) provide the numerical inputs for $\bar \alpha_j$.
They are determined from Eqs.~(\ref{3bAmp})~and~(\ref{RS0}), together with
the parameters in Eqs.~(\ref{coupling1})~and~(\ref{coupling2}),
and are given by
\begin{eqnarray}\label{alphavalue}
&&(\alpha_1,\alpha_2)=(0.17\pm 0.09, 0.33\pm 0.03)\,,\nonumber\\
&&(\alpha_3,\alpha_4)=(0.02\pm 0.01, 0.04\pm 0.01)\,,\nonumber\\
&&(\alpha_5,\alpha_6)=(0.59\pm 0.06,0.26\pm 0.03)\,.
\end{eqnarray}

To proceed, we perform a minimum $\chi^2$ fit defined as
\begin{eqnarray}\label{chi2}
\chi^2=\Sigma_i \bigg(\frac{{\cal B}_{\rm th}^i-{\cal B}_{\rm ex}^i}{\sigma_{\rm ex}^{i}}\bigg)^2
+\Sigma_j \bigg(\frac{\bar\alpha_j-\alpha_j}{\sigma_{\alpha_j}}\bigg)^2\,,
\end{eqnarray}
where the experimental branching fractions ${\cal B}_{\rm ex}^i$ in Eq.~(\ref{chi2})
and their uncertainties $\sigma_{\rm ex}^i$ are taken from Eq.~(\ref{6data}).
The quantities ${\cal B}_{\rm th}^i$ denote the corresponding theoretical branching fractions, 
which are evaluated as
${\cal B}_{\rm th}=({G_F^2|\vec{p}_{{\bf B}}|\tau_{\Lambda_c})/(16\pi m_{\Lambda_c}^2})
|{\cal M}(\Lambda_c^+\to{\bf B}S)|^2$. Here,
${\cal M}(\Lambda_c^+\to{\bf B}S)$ is expressed in terms of
the topological amplitudes listed in Tables~\ref{tab1} and \ref{tab2}, while
$\tau_{\Lambda_c}$ and $m_{\Lambda_c}$ denote the lifetime and mass of $\Lambda_c^+$,
respectively. The magnitude of the three-momentum 
for the final-state baryon in the $\Lambda_c^+$ rest frame
is given by
$|\vec{p}_{{\bf B}}|=[(m_{\Lambda_c}^2-M_+^2)(m_{\Lambda_c}^2-M_-^2)]^{1/2}/(2 m_{\Lambda_c})$,
with $M_\pm=m_{{\bf B}}\pm m_S$.
In Eq.~(\ref{chi2}), the coefficients $\bar\alpha_j$ are treated as 
fit parameters rather than fixed inputs and constrained by the second term 
in the $\chi^2$ function, while $\alpha_j$ and $\sigma_{\alpha_j}$ are taken from Eq.~(\ref{alphavalue}).
This treatment allows the theoretical resonant branching fractions 
${\cal B}_{\rm th(res)}$ in Eq.~(\ref{3bB}) to be more consistently matched to the
experimentally extracted resonant observables ${\cal B}_{\rm ex(res)}$ in Eq.~(\ref{6data}).

%
\begin{table}[b]
\caption{Branching fractions of ${\bf B}_c\to{\bf B}S$ calculated within
the topological-diagram approach, where the uncertainties reflect
those propagated from the mixing angles in Eq.~(\ref{theta12}) and
the fitted topological parameters in Eq.~(\ref{fitresults}). In the data column,
${\cal B}_{\rm ex(res)}(f_0,a_0)$ and ${\cal B}_{\rm ex(res)}(f_0,\sigma_0)$,
extracted in Eq.~(\ref{6data}), incorporate the information on
${\cal B}(\Lambda_c^+\to{\Sigma^{+} a_0^{0}})$,
${\cal B}(\Lambda_c^+\to \Sigma^{+} f_0)$, and
${\cal B}(\Lambda_c^+\to \Sigma^{+} \sigma_0)$, while
${\cal B}_{\rm ex(res)}(a_0^+)$ and ${\cal B}_{\rm ex(res)}(\kappa^+)$ correspond to
${\cal B}(\Lambda_c^+\to{\Sigma^{0} a_0^{+}})$ and
${\cal B}(\Lambda_c^+\to{\Xi^{0} \kappa^{+}})$, respectively.}\label{tab3}
{
\scriptsize
\begin{tabular}{|l|c|c|}
\hline
Branching fraction
&This work: $(S1,\,S2)$ &Data
\\
\hline\hline
$10^2{\cal B}(\Lambda_c^+\to{\Lambda^{0} a_0^{+}})$
&$(1.1\pm 0.4 ,1.1 \pm0.6)$
&$1.23\pm 0.21$~\cite{BESIII:2024mbf}\\
$10^2{\cal B}(\Lambda_c^+\to{\Sigma^{0} a_0^{+}})$
&$(1.3\pm0.8,1.2\pm 1.0)$
&${\cal B}_{\rm ex(res)}(a_0^+)$\\
$10^2{\cal B}(\Lambda_c^+\to{\Sigma^{+} a_0^{0}})$
&$(1.3\pm0.8,1.2\pm 1.0)$
&${\cal B}_{\rm ex(res)}(f_0,a_0)$\\
$10^2{\cal B}(\Lambda_c^+\to{\Xi^{0} \kappa^{+}})$
&$(1.4\pm 0.8,1.5 \pm 0.8)$
&${\cal B}_{\rm ex(res)}(\kappa^+)$\\
$10^2{\cal B}(\Lambda_c^+\to{p \bar{\kappa}^{0}})$
&$(0.21\pm 0.06,0.21\pm 0.06)$
&$0.19\pm0.06$~\cite{LHCb:2022sck,pdg}\\
$10^2{\cal B}(\Lambda_c^+\to \Sigma^{+} f_0)$
&$(0.5\pm 0.4,4.9\pm 1.9)$
&${\cal B}_{\rm ex(res)}(f_0,a_0)$, ${\cal B}_{\rm ex(res)}(f_0,\sigma_0)$\\
$10^2{\cal B}(\Lambda_c^+\to \Sigma^{+} \sigma_0)$
&$(3.4\pm1.7,0.1\pm 0.2)$
&${\cal B}_{\rm ex(res)}(f_0,\sigma_0)$\\\hline
$10^3{\cal B}(\Lambda_c^+\to{\Lambda^{0} \kappa^{+}})$
&$(1.2\pm 0.6,2.5\pm 1.0)$
&\\
$10^3{\cal B}(\Lambda_c^+\to{\Sigma^{0} \kappa^{+}})$
&$(0.6\pm 0.5,1.4\pm 0.8)$
&\\
$10^3{\cal B}(\Lambda_c^+\to{\Sigma^{+} \kappa^{0}})$
&$(2.5\pm 1.2,4.7\pm 2.0)$
&\\
$10^3{\cal B}(\Lambda_c^+\to{n a_0^{+}})$
&$(3.2\pm 1.5,3.1\pm 1.9)$
&\\
$10^3{\cal B}(\Lambda_c^+\to p a_0^{0})$
&$(0.7\pm 0.6, 0.7\pm 0.7)$
&\\
$10^3{\cal B}(\Lambda_c^+\to p f_0)$
&$(0.2\pm 0.1,3.6\pm1.4)$
&$3.4\pm 2.3$~\cite{Barlag:1990yv,pdg}
\\
$10^3{\cal B}(\Lambda_c^+\to p \sigma_0)$
&$(2.0\pm 0.9,0.05\pm 0.02)$
&\\\hline
$10^4{\cal B}(\Lambda_c^+\to{p \kappa^{0}})$
&$(1.2\pm 0.7,2.6\pm 1.2)$
&\\
$10^4{\cal B}(\Lambda_c^+\to{n \kappa^{+}})$
&$(1.2\pm 0.7,2.6\pm 1.2)$
&\\
\hline
\end{tabular}}
\end{table}
%

%
Regarding the experimental inputs, a residual correlation may arise between
${\cal B}_{\rm ex(res)}(f_0,\sigma_0)$ and ${\cal B}_{\rm ex(res)}(f_0,a_0)$,
as both observables depend on the same underlying branching fraction
${\cal B}(\Lambda_c^+\to \Sigma^+ f_0)$. Such a correlation could, in principle, 
influence the fit results. To assess this effect, we estimate the corresponding covariance 
and find that the resulting correlation is negligible.
Furthermore, since a standard $\chi^2$-fit procedure is adopted,
treating all experimental data as statistically independent inputs,
any residual correlation does not affect the individual central values or uncertainties. 
Consequently, even when this potential correlation is taken into account, 
the global fits remain robust and reliable.

The minimum $\chi^2$ fit, including the extraction of the parameter uncertainties,
is performed using the MINUIT2 package from the CERN ROOT framework~\cite{James:1975dr},
with the MIGRAD algorithm employed as the minimizer. This fitting procedure
has been widely used in phenomenological analyses, 
and its robustness and reliability are well established.
Two scenarios are considered in the global fit: S1 and S2, 
corresponding to the scalar configurations, 
$S_{q\bar q}$ in Eq.~(\ref{M_2q}) and $S_{q^2\bar q^2}$ in Eq.~(\ref{M_4q}),
respectively. The extracted topological parameters are
\begin{eqnarray}\label{fitresults}
{\rm S1}:\;&&
(|C'|, |E'|, |E_{{\bf B}}|)=(0.67\pm 0.16, 0.53\pm 0.13,0.15\pm 0.02)~{\rm GeV}^3\,,\nonumber\\
&&(\delta_{E'},\delta_{E_{{\bf B}}})=(-90.3\pm15.0, -144.4\pm40.7)^\circ\,,\nonumber\\
{\rm S2}:\;&&
(|C'|, |E'|, |E_{{\bf B}}|)=(0.97\pm 0.20,0.54\pm 0.13,0.15\pm 0.02)~{\rm GeV}^3\,,\nonumber\\
&&(\delta_{E'},\delta_{E_{{\bf B}}})=(-61.0\pm 17.7, -144.2\pm 39.6)^\circ\,.
\end{eqnarray}
The goodness of fit is quantified by $\chi^2/{\rm n.d.f.}$=2.6 for scenario S1 and 
0.6 for scenario S2, with the number of degrees of freedom fixed to n.d.f.=2 in both cases.
Using the fitted parameters in Eq.~(\ref{fitresults}),
the predicted branching fractions ${\cal B}(\Lambda_c^+ \to {\bf B}S)$
are presented in Table~\ref{tab3}.

\section{Discussions and Conclusion}
Within the $SU(3)_f$ framework, the measured branching fractions of $\Lambda_c^+\to{\bf B}S$
listed in Eq.~(\ref{6data}) are subjected to a global fit, in which a common set of 
$SU(3)_f$ topological parameters is employed to relate the corresponding decay amplitudes.
As shown in Eq.~(\ref{fitresults}), the two sets of extracted values 
for $|E'|$, $|E_{\bf B}|$, and $\delta_{E_{\bf B}}$ are nearly identical, while 
those for $|C'|$ and $\delta_{E'}$ remain mutually consistent once uncertainties 
are taken into account. Nevertheless, 
the overall fit quality in scenario S1 is notably poorer than that in scenario S2,
as quantified by the corresponding $\chi^2/n.d.f.$ values.

The relatively poor description in scenario S1 arises because the amplitudes
${\cal M}_{q\bar q}(\Lambda_c^+ \to \Sigma^+ f_0)\simeq E_{\bf B} c\phi_I$ and
${\cal M}_{q\bar q}(\Lambda_c^+ \to p f_0)\simeq \frac{1}{\sqrt 2}(C^\prime+E^\prime)s\phi_I$,
together with the small magnitudes of both $E_{\bf B}$ and $s\phi_I$,
lead to theoretical branching fractions that underestimate 
for the measured values ${\cal B}_{\rm ex(res)}(f_0,a_0)$ and ${\cal B}_{\rm ex}(p f_0)$, respectively.
Consequently, the fit yields $\chi^2/{\rm n.d.f.}=2.6$, which deviates significantly from unity.
In contrast, scenario S2 achieves a substantially improved fit quality with $\chi^2/n.d.f.=0.6$.
In this case, ${\cal M}_{q^2\bar q^2}(\Lambda_c^+ \to \Sigma^+ f_0,p f_0)\simeq
\frac{1}{\sqrt 2}(C^\prime+E^\prime)c\phi_{II}$ lead to significantly enhanced branching fractions,
consistent with the hierarchy $c\phi_{II}\simeq 1 \gg s\phi_{II}\simeq 0$.
This improvement suggests that the tetraquark interpretation of the light scalar mesons 
provides a more consistent description of the current experimental data within the present framework.

Nonetheless, we obtain the relation ${\cal B}_{q\bar q}\simeq {\cal B}_{q^2\bar q^2}$
for most decay modes, indicating that the predicted branching fractions of $\Lambda_c^+\to{\bf B}S$ 
are largely insensitive to the internal structure of the light scalars,
except for channels involving $f_0-\sigma_0$ mixing. 
To further probe the nature of the light scalar in $\Lambda_c^+\to{\bf B}S$,
we identify the following relations
\begin{eqnarray}\label{R1R2}
&&
\frac{{\cal B}_{q\bar q}(\Lambda_c^+\to p f_0)}{{\cal B}_{q\bar q}(\Lambda_c^+\to p \sigma_0)}
=\tan^2\theta_{I}\times (\Phi_{pf_0}/\Phi_{p\sigma_0})=0.127\,,\nonumber\\
&&
\frac{{\cal B}_{q^2\bar q^2}(\Lambda_c^+\to p \sigma_0)}{{\cal B}_{q^2\bar q^2}(\Lambda_c^+\to p f_0)}
=\tan^2\theta_{II}\times (\Phi_{p\sigma_0}/\Phi_{pf_0})=0.015\,,
\end{eqnarray}
where $\tan^2\theta_I=0.19$, $\tan^2\theta_{II}=0.01$, and $(\Phi_{f_0}/\Phi_{\sigma_0})=0.67$ 
denotes the ratio of the integrated phase space factors defined below Eq.~(\ref{chi2}). Importantly,
these ratios are independent of the topological parameters and their uncertainties.
Consequently, future experimental measurements can directly test the contrasting predictions
${\cal B}_{q\bar q}(\Lambda_c^+\to p f_0)=0.127{\cal B}_{q\bar q}(\Lambda_c^+\to p\sigma_0)$ and
${\cal B}_{q^2\bar q^2}(\Lambda_c^+\to p\sigma_0)=0.015{\cal B}_{q^2\bar q^2}(\Lambda_c^+\to p f_0)$,
which may provide a clear discriminator between the two scenarios. Similarly,
the relations ${\cal B}_{q\bar q}(\Lambda_c^+\to \Sigma^+ f_0)
\simeq 0.15{\cal B}_{q\bar q}(\Lambda_c^+\to \Sigma^+\sigma_0)$ and 
${\cal B}_{q^2\bar q^2}(\Lambda_c^+\to \Sigma^+\sigma_0)
\simeq 0.02{\cal B}_{q^2\bar q^2}(\Lambda_c^+\to \Sigma^+ f_0)$ 
offer additional observables with comparable discriminating power.

With the factorizable $W$-emission topologies being negligible ($T\simeq C\simeq 0$),
the decays proceed exclusively through $W$-exchange contributions. In contrast, 
the decays $\Lambda_c^+\to{\bf B}M$ can receive factorizable contributions 
from $W$-emission diagrams. Nonetheless,
we predict ${\cal B}(\Lambda_c^+\to \Sigma^{0(+)} a_0^{+(0)})\simeq
{\cal B}(\Lambda_c^+\to \Lambda a_0^+)\simeq 1\times 10^{-2}$, 
which are comparable in magnitude to the experimental measured branching fractions
${\cal B}(\Lambda_c^+\to \Sigma^{0(+)} \pi^{+(0)})\simeq 
{\cal B}(\Lambda_c^+\to \Lambda \pi^+)\simeq 1 \times 10^{-2}$ reported in Ref.~\cite{pdg}. 
We further find that
${\cal B}(\Lambda_c^+\to\Xi^{0} \kappa^{+})\simeq 
{\cal B}(\Lambda_c^+\to \Xi^{0} K^{+})$~\cite{BESIII:2018cvs}, 
with both channels receiving contributions solely from the $E^{\prime}$ topology.
More generally, the CA and SCS branching fractions of $\Lambda_c^+\to {\bf B}S$  
are typically at the level of $10^{-2}$ and $10^{-3}$, and are
compatible with the corresponding CA and SCS branching fractions 
of $\Lambda_c^+\to {\bf B}M$, respectively. This consistency
supports the conclusion that, within the short-distance picture,
${\bf B}_c\to {\bf B}S$ decays can naturally attain branching fractions comparable to those of
${\bf B}_c\to {\bf B}M$, rendering these channels promising targets 
for experimental investigation in the near future.

The $SU(3)_f$ approach rests on the assumption that the light quarks have nearly equal masses.
In reality, the pronounced mass hierarchy $m_s \gg m_{u,d}$ inevitably leads to $SU(3)_f$ symmetry breaking.
As discussed in Ref.~\cite{Gronau:1995hm}, within the topological-diagrammatic framework 
such breaking effects can arise naturally from flavor flows involving the s quark.
Empirically, observables such as
${\cal B}(D^0\to K^+K^-)/{\cal B}(D^0\to \pi^+\pi^-)$~\cite{Li:2012cfa,Cheng:2019ggx},
${\cal B}(\Lambda_c^+ \to \Sigma(1385)^+ \eta)$~\cite{Hsiao:2020iwc}, and
${\cal B}(\Xi_c^0\to \Xi^- K^+)/{\cal B}(\Xi_c^0\to \Xi^- \pi^+)$~\cite{Hsiao:2021nsc}
consistently indicate that the $g\to s\bar s$ mechanism gives rise to
the most significant sources of $SU(3)_f$-breaking effects. 

Accordingly,
in Tables~\ref{tab1} and~\ref{tab2}, we introduce the $W$-exchange topologies 
$E_{\bf B}^{s}$, $E^{\prime s}$, and $E_S^{\prime s}$ 
as a basis for discussing the $SU(3)_f$ symmetry breaking induced by $g\to s\bar s$.
To quantify these effects, we further parameterize
$E_{\bf B}^{s}=E_{\bf B}+\delta E_{\bf B}^s$,
$E^{\prime s}=E^{\prime}+\delta E^{\prime s}$, and
$E_S^{\prime s}=E_S^{\prime}+\delta E_S^{\prime s}$ following Ref.~\cite{Wang:2023uea},
where $\delta E_{\bf B}^s$, $\delta E^{\prime s}$, and $\delta E_S^{\prime s}$
represent the corresponding breaking contributions.
This leads to the approximate relations for the $\kappa$ modes,
\begin{eqnarray}
&&
{\cal B}_{\rm bk}(\Lambda_c^+\to \Sigma^0\kappa^+)
\simeq(1+\delta E_{\bf B}^s/E_{\bf B})^2 {\cal B}_{\rm th}(\Lambda_c^+\to \Sigma^0\kappa^+)\,,
\nonumber\\
&&
{\cal B}_{\rm bk}(\Lambda_c^+\to \Sigma^+\kappa^0)
\simeq(1+\delta E^{\prime s}_S/E'_S)^2 {\cal B}_{\rm th}(\Lambda_c^+\to \Sigma^+\kappa^0)\,,
\nonumber\\
&&
{\cal B}_{\rm bk}(\Lambda_c^+\to \Xi^0\kappa^+)
\simeq(1+\delta E^{\prime s}/E^{\prime})^2 {\cal B}_{\rm th}(\Lambda_c^+\to \Xi^0\kappa^+)\,,
\end{eqnarray}
where ${\cal B}_{\rm bk}$ denotes the branching fraction including $SU(3)_f$ breaking effects.
Consequently, any deviation of ${\cal B}_{\rm bk}$ from the symmetric prediction ${\cal B}_{\rm th}$
provides a direct probe of flavor-symmetry breaking in future measurements.
Given that the $SU(3)_f$-breaking effects may reach the level of
40-70\% ({\it i.e.}, $\delta E^s/E\simeq 0.4-0.7$)~\cite{Li:2012cfa,Cheng:2019ggx,
Hsiao:2020iwc,Hsiao:2021nsc}, one may therefore expect experimentally 
observable enhancements as large as ${\cal B}_{\rm bk}\simeq 2{\cal B}_{\rm th}$.

In summary, we have demonstrated that short-distance contributions, described by
the topological diagrams within the $SU(3)_f$ framework, play a dominant role in ${\bf B}_c\to{\bf B}S$.
Using the currently available experimental data, we extracted the relevant topological amplitudes
and applied them to compute branching fractions across various decay channels.
Under the assumption that light scalar mesons are conventional $q\bar q$ states,
the global fit yields $\chi^2/n.d.f=2.6$, significantly away from unity.  In contrast,  
the $q^2\bar q^2$ picture is more strongly favored by the present measurements.
Within this framework, our analysis successfully accounted for the observed branching fraction of
$\Lambda_c^+\to \Lambda a_0^+$, which is an order of magnitude
larger than predictions based solely on long-distance effects.
Treating the light scalar mesons as tetraquark candidates, we predicted
${\cal B}_{q^2\bar q^2}(\Lambda_c^+\to \Sigma^+ f_0)=(4.9\pm 1.9)\times 10^{-2}$ and
${\cal B}_{q^2\bar q^2}(\Lambda_c^+\to p f_0)=(3.6\pm 1.4)\times 10^{-3}$, while
${\cal B}_{q^2\bar q^2}(\Lambda_c^+\to \Sigma^+ \sigma_0)$ and 
${\cal B}_{q^2\bar q^2}(\Lambda_c^+\to p\sigma_0)$
are suppressed to the levels of $1\times 10^{-3}$ and $5\times 10^{-5}$, respectively.
Moreover, the decay $\Lambda_c^+\to\Xi^{0} \kappa^{+}$,
which receives a sole contribution from a non-factorizable $W$-exchange diagram,
is predicted to have
${\cal B}_{q^2\bar q^2}(\Lambda_c^+\to\Xi^{0} \kappa^{+})=(1.5 \pm 0.8)\times 10^{-2}$.
More generally, the branching fractions of $\Lambda_c^+\to{\bf B}S$ 
are found to be comparable to those of $\Lambda_c^+\to{\bf B}M$,
indicating that these channels are promising targets 
for experimental studies at BESIII, Belle~II, and LHCb.

\section*{ACKNOWLEDGMENTS}
This work was supported in part by National Natural Science Foundation of China
(Grants~No.~12575101 and No.~12175128).

\end{document}